\documentclass[12pt,preprint]{aastex}
\usepackage{emulateapj5}
\submitted{To appear in the Astrophysical Journal}

\shorttitle{OPTICAL SPECTRAL MONITORING OF XTE~J1118+480 IN OUTBURST}
\shortauthors{TORRES ET AL.}

\begin{document}

\title{Optical Spectral Monitoring of XTE~J1118+480 in Outburst: Evidence for a Precessing Accretion Disk?}
\author{M. A. P. Torres\altaffilmark{1}, P. J. Callanan\altaffilmark{1}, M. R. Garcia\altaffilmark{2}, J. E. McClintock\altaffilmark{2}, P. Garnavich\altaffilmark{3}, Z. Balog\altaffilmark{2}\altaffilmark{,4}, P. Berlind\altaffilmark{2}, W. R. Brown\altaffilmark{2}, M. Calkins\altaffilmark{2}, A. Mahdavi\altaffilmark{2}}

\altaffiltext{1}{Physics Department, University College, Cork, Ireland; mapt@phys.ucc.ie, paulc@ucc.ie}

\altaffiltext{2}{Harvard-Smithsonian Center for Astrophysics, 60 Garden St, Cambridge, MA 02138; mgarcia@cfa.harvard.edu, jem@cfa.harvard.edu, pzhao@cfa.harvard.edu, pberlind@cfa.harvard.edu, wbrown@cfa.harvard.edu, mcalkins@cfa.harvard.edu, amahdavi@cfa.harvard.edu}

\altaffiltext{3}{Department of Physics. University of Notre Dame, 213 Nieuwland Science Hall, Notre Dame, Indiana 46556-5670; pgarnavi@nd.edu}     

\altaffiltext{4}{on leave from Dept of Optics and Quantum Electronics, University of Szeged, Dom ter 9., H-6720 Szeged, Hungary}  
\begin{abstract} 

We present spectroscopic observations of the X-ray transient XTE~J1118+480 acquired during different epochs following the 2000 March outburst. We find that the emission line profiles show variations in their double-peak structure on time scales longer than the 4.1 hr orbital period. We suggest that these changes are due to a tidally driven precessing disk. Doppler imaging of the more intense Balmer lines and the He{\sc ii} $\lambda$4686 line shows evidence of a persistent region of enhanced intensity superposed on the disk which is probably associated with the gas stream, the hotspot or both. We discuss the possible origins of the optical flux in the system and conclude that it may be due to a viscously heated disk.

\end{abstract} 


\section{INTRODUCTION}

X-ray Novae (XRNe) form a subclass of Low Mass X-ray binaries (LMXBs) which has provided us with some of the best stellar-mass black hole candidates (see the reviews by \citealt{van95}, \citealt{van98} and \citealt{cha98}). These systems undergo occasional outbursts, with a recurrence time scale of decades, caused by an episode of intense mass transfer onto the compact primary via an accretion disk. A disk instability \citep{can93,can98} is generally invoked as the triggering mechanism for the outburst. 

\begin{figure}
\epsscale{0.8}
\plotone{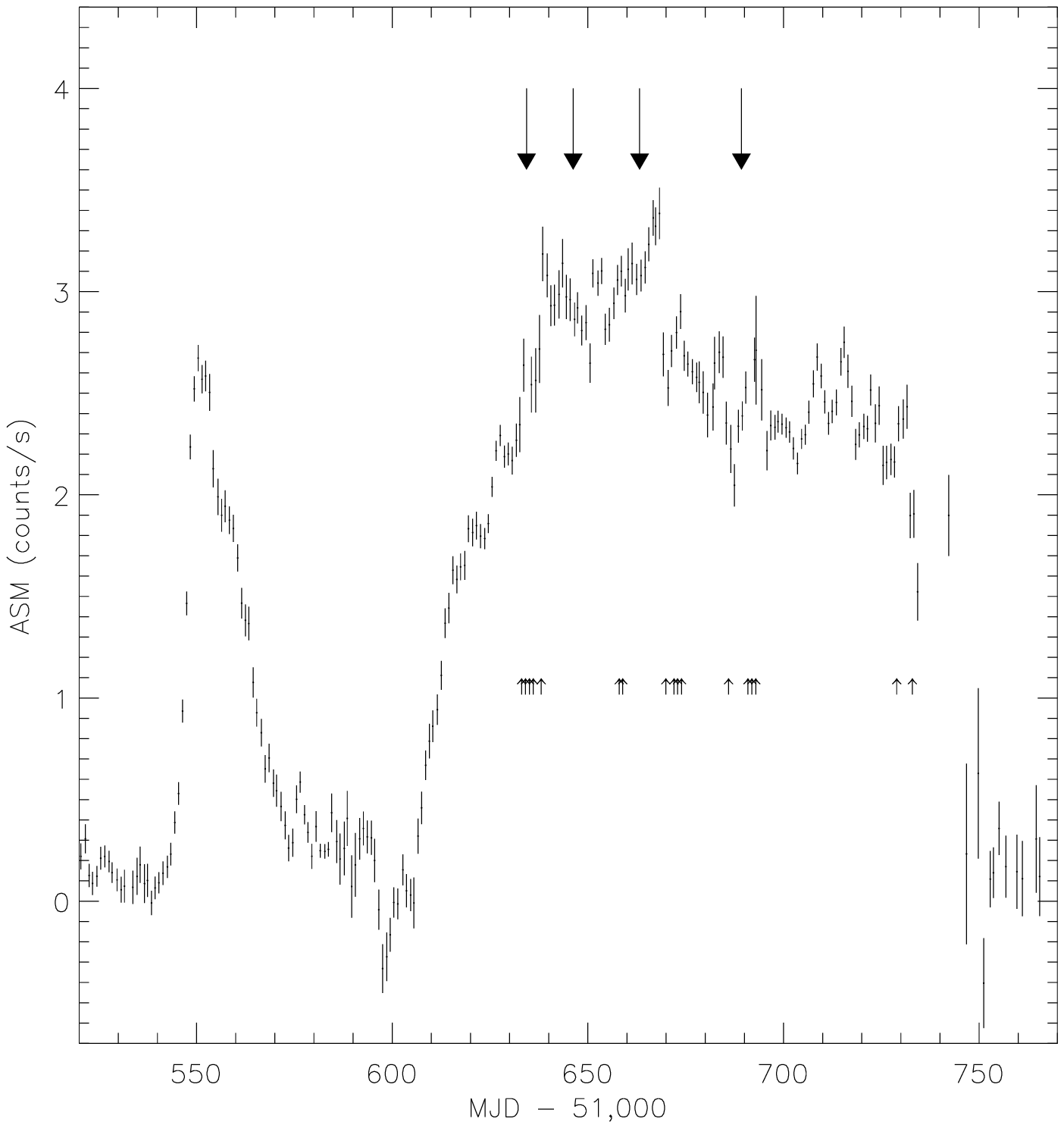}
\caption[f1.eps]{RXTE ASM outburst light curve of XTE J11184+480 after smoothing with a 3 point weighted average. The arrows indicate the dates of spectroscopic observations: long solid arrows mark the nights during which the orbital phase coverage allowed us to carry out Doppler tomography (see Table~\ref{Emissiontable}). \label{fig1}}
\end{figure}

The XRN XTE~J1118+480 was discovered on 2000 March 29 by the Rossi X-Ray Timing Explorer All-Sky Monitor (RXTE ASM) at the beginning of a prolonged outburst \citep{rem00}. Reanalysis of the ASM data showed a previous outburst episode during January 2-29 (MJDs~$51,545-51,572$; see Figure \ref{fig1}). The optical counterpart was identified with a 12.9 magnitude star coincident with an 18.8 magnitude object in the USNO catalogues \citep{uem00A}. \citet{uem00B} claimed the detection of superhumps in the outburst light curve with a periodicity of 0.1709~d. The spectrum of the optical counterpart was typical of X-ray binaries in outburst \citep{gar00}, but the X-ray flux was low for an X-ray transient \citep{rem00}. A high inclination system \citep{gar00} and/or a 'mini-outburst' state \citep{hyn00} have been suggested to explain the anomalous X-ray flux. \citet{dub01} presented the first constraints on the system parameters from the analysis of spectroscopic observations during the outburst. \citet{mcc00,mcc01A} and \citet{wag00,wag01} derived an orbital period that was shorter than the Uemura's determination (by $0.4-0.6~\%$ respectively) from radial velocity measurements near quiescence. These authors determined the mass of the primary star to be $>6~M_\odot$. This places XTE~J1118+480 among the dynamically confirmed black holes.

In this paper we present the results from a systematic spectroscopic campaign. Data were collected on 21 nights during 2000 March-July; on four of these nights we obtained almost complete orbital phase coverage.
 
\section{OBSERVATIONS AND DATA REDUCTION} 

We monitored XTE~J1118+480 for more than three months, beginning 2000 March 31 (UT) and continuing until July 10. The VSNET light curve, which covers most of this period, shows that the source brightness was V=13$\pm0.3$ during 2000 April 3 - June 6. Optical spectra were obtained using the FAST spectrograph \citep{fab98} attached to the 1.5~m Tillinghast telescope of the Fred Lawrence Whipple Observatory. The spectra cover the 3600 - 7500~\AA~range with a resolution of 4.5~\AA~FWHM. A HeNeAr lamp spectrum was taken inmediately after each object exposure in order to determine the wavelength calibration. The flux standards Feige 34, Feige 56, Hiltner 600, and BD+33 2642 were observed throughout the run in order to correct for the instrumental response. While occasional clouds and variable seeing do not allow an absolute flux calibration, the shape of the continuum is reliable except at the extreme blue and red ends (shortward of 4000~\AA~and longward of 6800 \AA). Spectra were extracted, wavelength- and flux-calibrated using the KPNO IRAF package. Details of the observations are summarized in Table~\ref{Emissiontable}.

\section{ANALYSIS}

\subsection{The Averaged Spectrum}

\begin{figure}
\epsscale{0.8}
\plotone{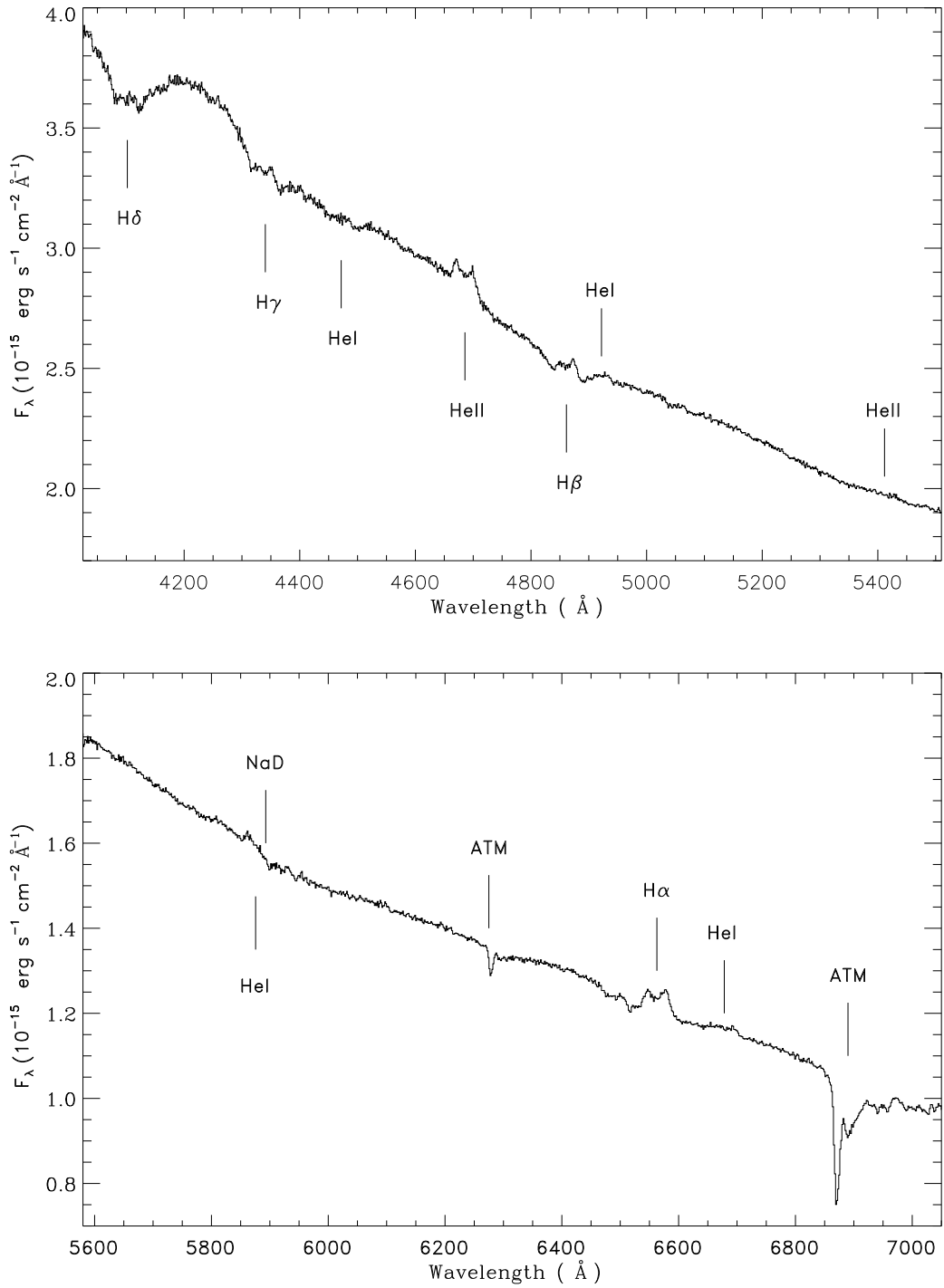}
\caption[f2.eps]{Average optical FAST spectrum of XTE J1118+480 on March 31 (UT). Major features are identified. ATM denotes the atmospheric features. \label{fig2}}
\end{figure}

In Figure~\ref{fig2}, we present the spectrum of XTE~J1118+480 obtained by averaging the individual spectra observed on night March 31. The spectrum shows clearly the presence of broad double-peaked emission lines of He{\sc ii} $\lambda$4686 and the Balmer series (up to H$\delta$). Weaker He{\sc ii} $\lambda5412$ and emission lines from He{\sc i} at $\lambda\lambda$4471, 4921, 5875, and 6678 are recognizable. Except for H$\alpha$, the emission cores in the Balmer lines and He{\sc i} $\lambda4471$ are contained within broad absorption features. The Bowen blend $\lambda\lambda 4640-4650$ is not present during this night (but see below). The broad bump at 4200 \AA~is probably an instrumental artifact. The main interstellar feature is the faint blend due to the Na D doublet at 5890 and 5896~\AA. We measure the equivalent width of this feature to be $\sim0.02$~\AA~(close to the noise level) indicating very low interstellar absorption. However, we cannot exclude a contribution from the He{\sc i} $\lambda$5875 emission line and consequently the measured equivalent width is an upper limit. Longward of $\sim$ 6800 \AA, the spectra are contaminated by telluric features. 

\begin{figure}
\epsscale{0.8}
\plotone{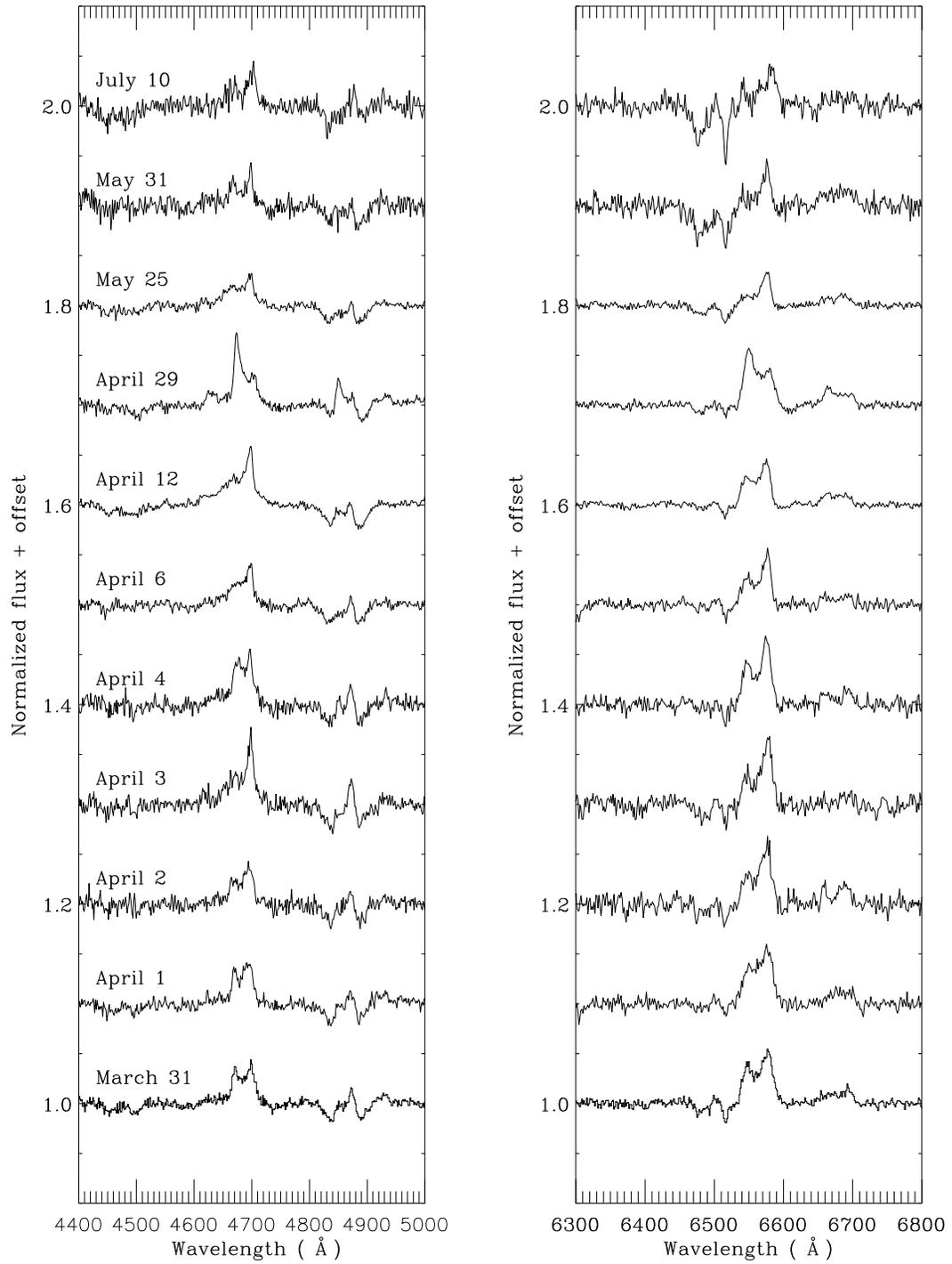}
\caption[f3.eps]{Evolution of the averaged red and blue spectrum during outburst. Individual spectra have been averaged on a night by night basis when more than two spectra were available.\label{fig3}}
\end{figure}

\begin{figure}
\epsscale{0.8}
\plotone{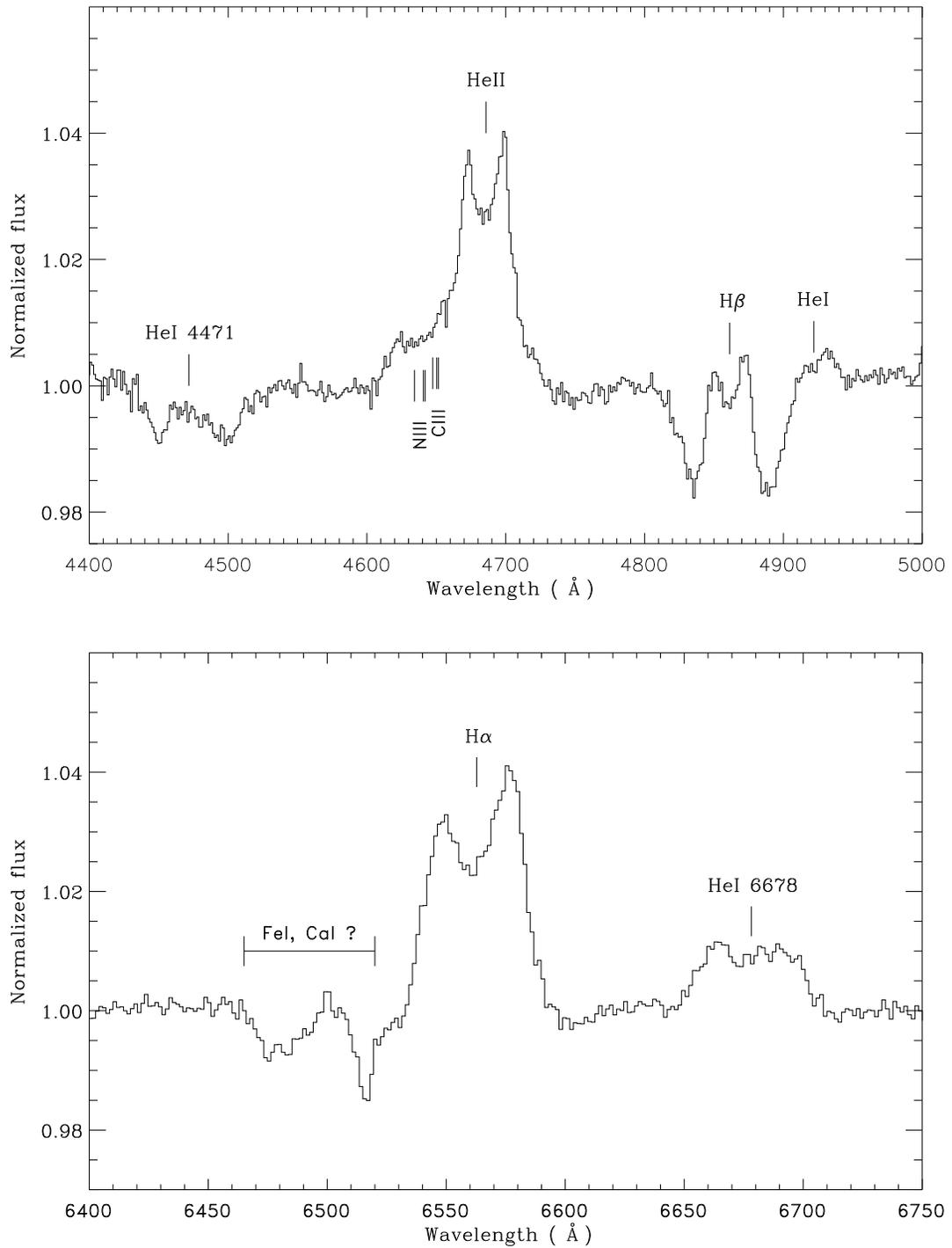}
\caption[f4.eps]{Average of the blue and red spectra of XTE J1118+480 acquired during our run. Note the two absorption features blueward H$\alpha$.\label{fig4}}
\end{figure}

As reported by \citet{gar00}, two unusual absorption features at 6479 and 6516~\AA~were present on the blueward side of the H$\alpha$ profile in all our observations (see Figures~\ref{fig3} and~\ref{fig4}). They must be associated with the accretion disk because in outburst the disk spectrum strongly dominates over the spectrum of the faint (V=18.8) secondary star. We suggest that both features are due to the superposition of a single broad absorption trough and an emission component due to a blend of Fe{\sc i}/Ca{\sc i} lines. This metallic blend is observed in absorption at $\lambda6495$ in late G-K stars \citep{hor86}, which implies that it originates in regions of the disk with T$_{eff}\sim$~4200-6000 K. It is also possible that Fe{\sc ii}~$\lambda6516$ contributes to the absorption features at this wavelength. 

The spectra acquired during the same night or contigous nights (when the number of spectra was limited) were averaged. The resulting mean spectra were de-reddened using $E(B-V)=0.024$ and $A{_V}/E(B-V)=3.1$. The colour excess was estimated by using the hydrogen column in the line of sight (N${_H}=1.34\times10^{20}$ cm$^{-2}$;~\citealt{dic90}) and the relation between N$_H$ and $E(B-V)$ of~\citet{boh78}. Next we fitted the de-reddened spectra with a power law of the form ${F_\lambda}\propto{\lambda^\alpha}$ after masking the major emission lines. We measure a mean power law index of $\alpha=-2.5\pm0.1$ in agreement with the values reported by \citet{dub01}.

\subsection{The Long-term Evolution of the Emission Lines}

We studied the more intense emission lines as follows: for each spectrum we fitted a low order spline to the adjacent continuum of every profile after masking the lines, and the spectrum was divided by the fitted function. Next, for those nights during which we obtained more than two spectra, we averaged the spectra. The H$\alpha$ and He{\sc ii} $\lambda$4686 line profiles were fitted with a 2-gaussian function and the H$\beta$ profile with a 3-gaussian function (with one used to account for the absorption component) using the Marquardt algorithm \citep{bev69}. Tables~\ref{haheii} and \ref{hb} list the values of the main fitted line parameters for the nights with good orbital coverage. In Table~\ref{fwzi} we give the measured full width zero intensity (FWZI) and equivalent widths (EWs) for H$\alpha$, H$\beta$, and He{\sc ii} $\lambda$4686. The EWs for additional lines are presented in Table~\ref{ew}. The uncertainties in the EWs and FWZIs were estimated by looking at the scatter in the values when selecting different wavelength intervals to set the local continuum level.

The mean peak-to-peak velocity separation for the H$\alpha$ and H$\beta$ lines is comparable ($\sim$~1200 km s$^{-1}$) with little change between different epochs. The peak-to-peak velocity separation of the He{\sc ii} $\lambda$4686 line ranges from 1600 to 2000 km s$^{-1}$. This result suggests that the emission at H$\alpha$ and H$\beta$ arises at a similar distance from the compact object and the He{\sc ii} $\lambda$4686 line is emitted from regions closer to the compact object. The velocity separation of the double-peaks in the H$\alpha$ line is consistent with the values reported for other black hole XRNe during quiescence and outburst (see e.g. Table 1 in \citealt{smi99}). The FWZI of the H$\beta$ absorption component implies a projected velocity of $\sim$~2900 km s$^{-1}$ for the inner part of the optically-thick accretion disk emitting at this wavelength.

The most remarkable feature in the evolution of the emission lines is the change in the double-peak intensity, from almost symmetric peaks to enhanced redshifted or blueshifted peaks (see Figure~\ref{fig3}). Even though this behavior in the poorly sampled nights might be attributable to an S-wave, a different explanation is required for the symmetric/asymmetric averaged profiles on March 31, April 12, 29, and May 25, where uniform orbital phase coverage should ensure the cancellation of any S-wave effect in the averaged profiles. 

Finally, the Bowen blend appears marginally on April 3, probably blended with He{\sc ii} $\lambda$4686 on April 12 and clearly enhanced on April 29. A glance at the RXTE/ASM light curve (1.3-12.2 keV) of XTE~J1118+480 (see Figure~\ref{fig1}) shows that the enhancement in intensity occurs near the maximum in X-ray flux. On May 25, during the slow decay of the X-ray outburst, both the Bowen Blend and He{\sc ii} $\lambda$4686 have decreased in intensity. 

\section{DOPPLER TOMOGRAPHY}\label{tomography}

Although the emission lines show variations in the double-peak structure due to an S-wave emission component, our attempts to find a periodic radial velocity modulation failed due mainly to our low spectral resolution and the complexity of the line profiles. Fortunately, we can obtain valuable information by using the Doppler Tomography technique \citep{mar88} on those data sets with good orbital phase coverage. This technique reconstructs the brightness distribution of the binary system in velocity space, allowing us to localize the emission structures which are not easily recognizable in the individual spectra.

We use the maximum-entropy method (MEM) of building the tomograms, which gradually builds the emission structure from a default uniform image by reducing the $\chi^2$ between the data and the model fit. The optimal solution is selected among infinite possibilities by maximizing the entropy of the image. To obtain the tomograms, the orbital phases were determined using an orbital period P${_{orb}}=0.169937\pm0.000007$~d and a time of minimum light T$_o$(HJD)$=2,451,880.1060\pm0.0011$~\footnotemark[5], which corresponds to closest approach of the secondary to the observer.

\footnotetext[5]{This orbital period was obtained by using V-band photometry acquired during 2001 January 20, 21 and February 16, 17, and 19 with the 1.2~m telescope at the F. L. Whipple Observatory. The time of minimum light was derived from the spectroscopic ephemeris given in \citet{mcc01A}}

\footnotetext[6]{This technique has the advantage that no treatment is necessary for negative values in the spectra.}

\begin{figure}
\begin{center}
\includegraphics[angle=-90,width=7.0in]{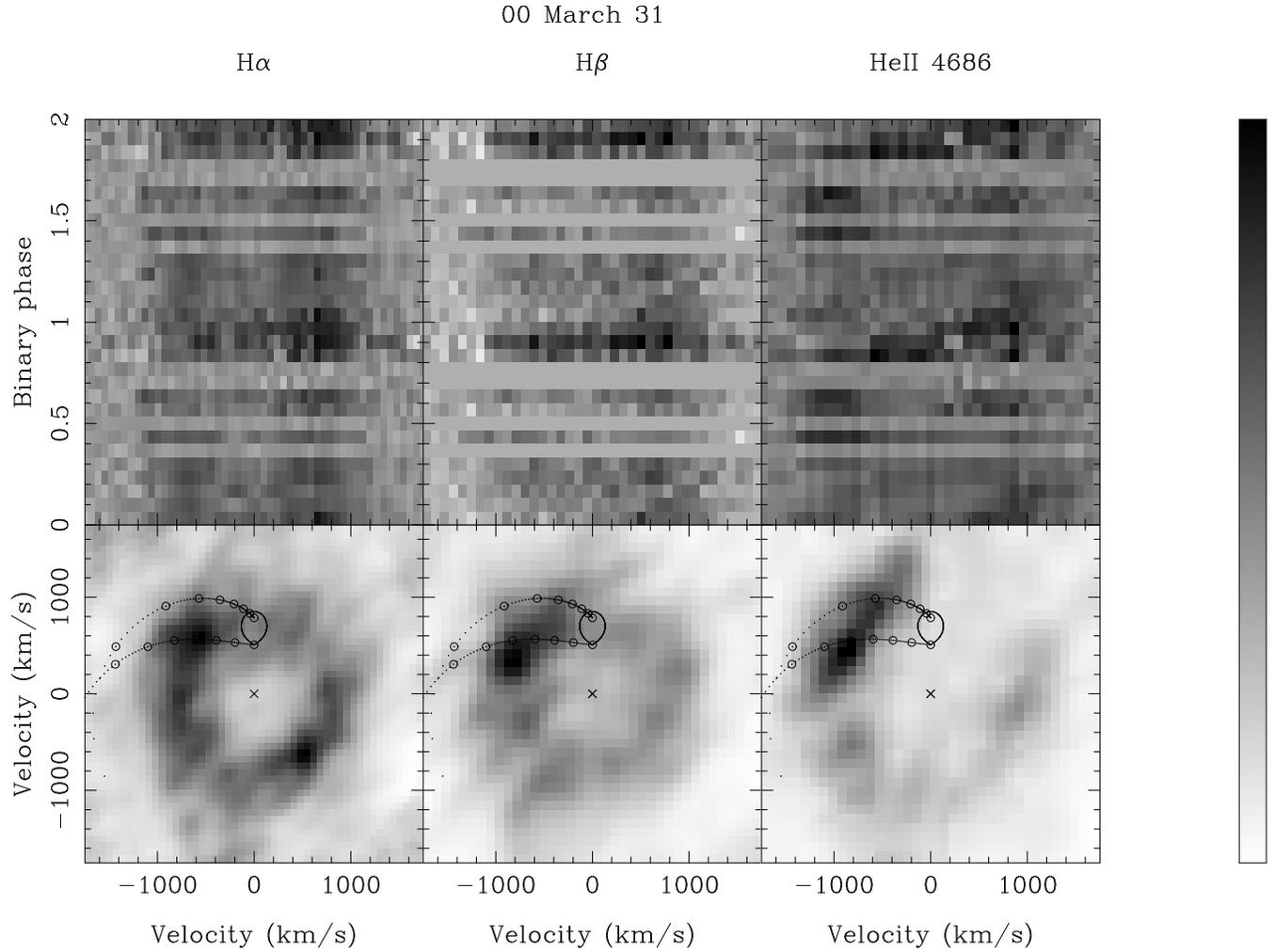}  
\caption[f5.eps]{Top panels present from left to right the trailed spectrograms of H$\alpha$, H$\beta$ and He{\sc ii}~$\lambda4686$. Empty strips represent gaps in the phase coverage. For the sake of clarity, the same cycle has been plotted twice. Lower panels show the computed MEM Doppler maps. The Roche lobe of the secondary star, the predicted velocities of the gas stream (lower curve) and of the disk along the stream (upper curve) are plotted for K${_2}$=698 km s$^{-1}$ and $q=0.07$. Distances in multiples of 0.1R$_{L1}$ are marked along both curves with open circles. The center of mass of the system is denoted by a cross.\label{fig5}}
\end{center}
\end{figure}

\begin{figure}
\begin{center}
\includegraphics[angle=-90,width=7.0in]{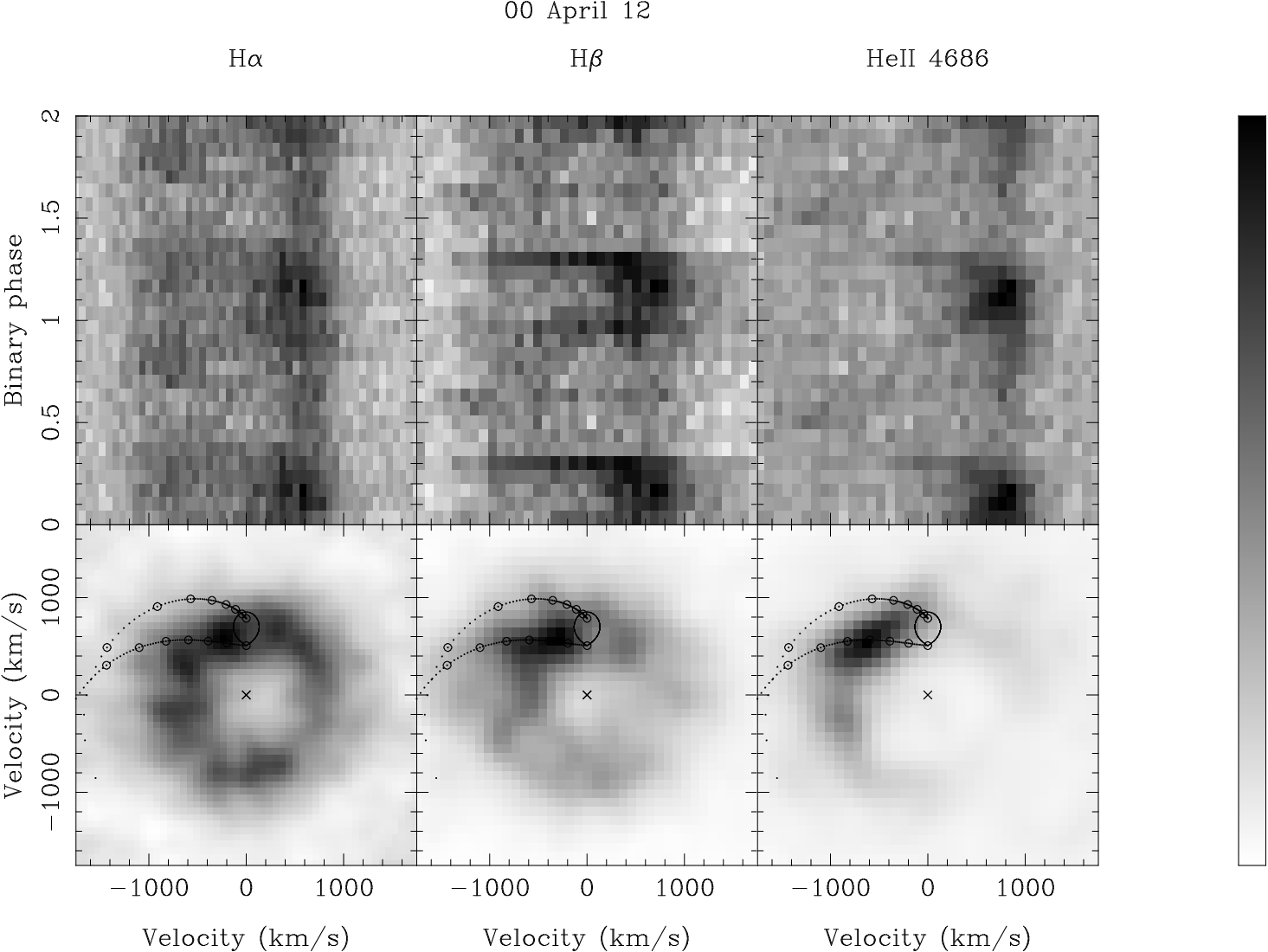}
\caption[f6.eps]{Trailed spectrogram and Doppler maps of H$\alpha$, H$\beta$ and He{\sc ii}~$\lambda4686$ built with data acquired during April 12.\label{fig6}}
\end{center}
\end{figure}

\begin{figure}
\begin{center}
\includegraphics[angle=-90,width=7.0in]{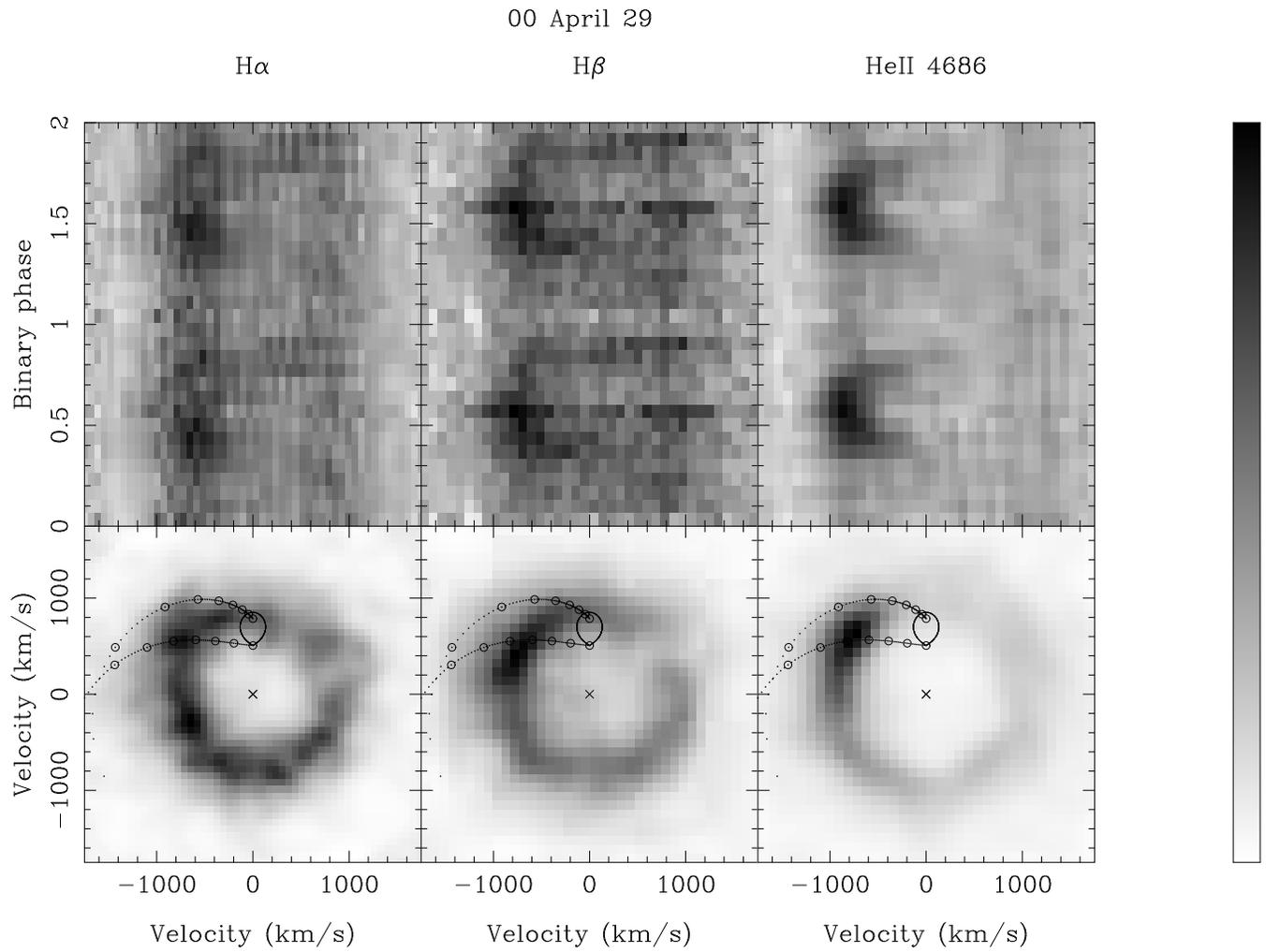}
\caption[f7.eps]{Trailed spectrogram and Doppler maps of H$\alpha$, H$\beta$ and He{\sc ii}~$\lambda4686$ built with data acquired during April 29.\label{fig7}}
\end{center}
\end{figure}

\begin{figure}
\begin{center}
\includegraphics[angle=-90,width=7.0in]{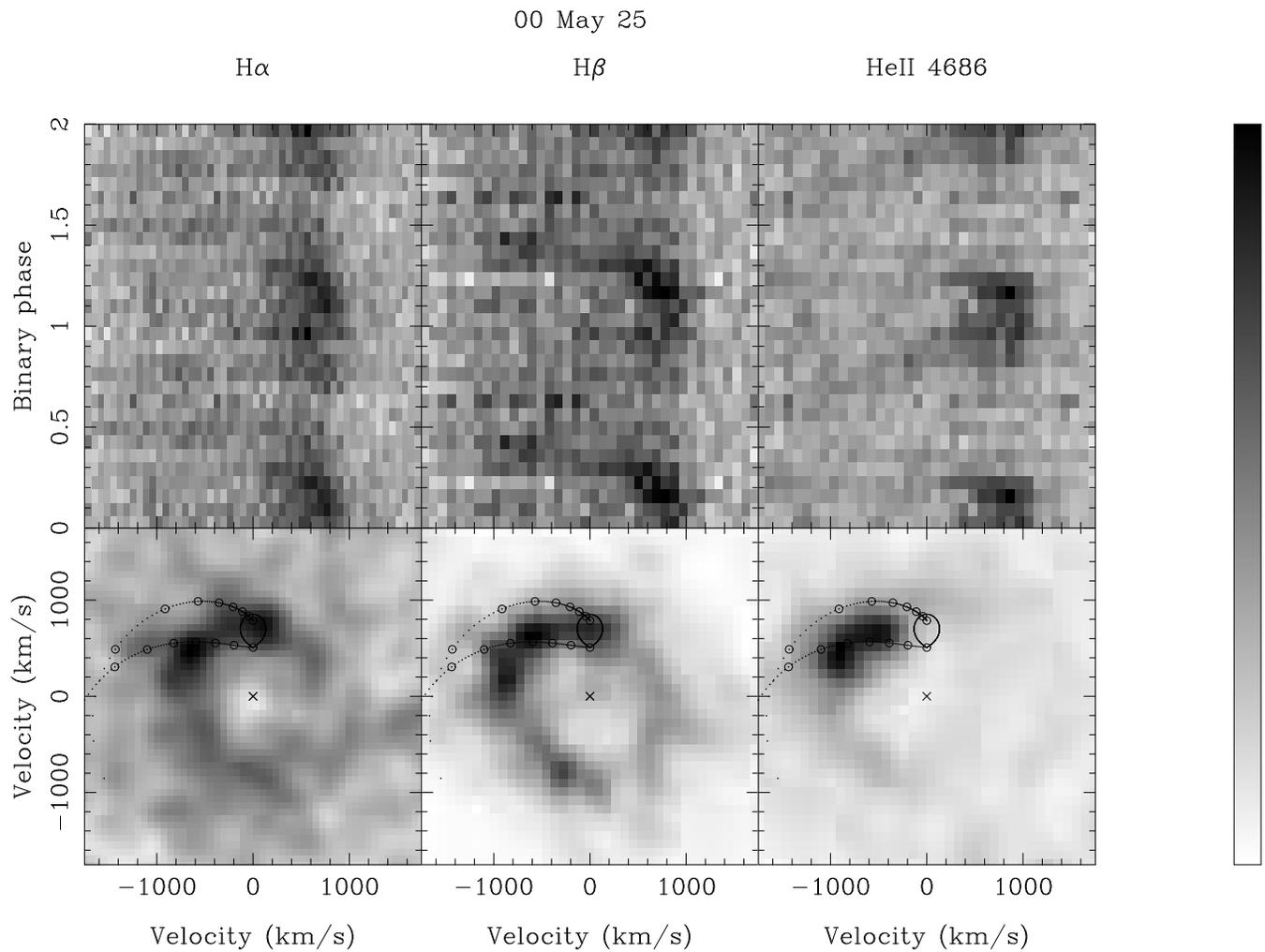}
\caption[f8.eps]{Trailed spectrogram and Doppler maps of H$\alpha$, H$\beta$ and He{\sc ii}~$\lambda4686$ built with data acquired during May 25. The H$\beta$ map suggests an eccentric disk. Note that only the FBP tomogram was computed for H$\alpha$ because of the proximity of the blueward absorption feature to the wing of the line (see Figure 3). \label{fig8}}
\end{center}
\end{figure}

We have computed Doppler maps of the He{\sc ii}~$\lambda$4686, H$\alpha$, and H$\beta$ lines for the nights March 31, April 12, 29, and May 25. Because MEM Doppler tomography cannot deal with negative values, the broad absorption in the H$\beta$ line was removed from each spectrum through a high-order spline fit to the profile, after masking the emission core. Figures~\ref{fig5}-\ref{fig8} display the derived Doppler maps: the upper panels show the spectra in the form of trailed spectrograms constructed by folding the line profile spectra into 15 phase bins. The bottom panels show the MEM tomograms. Similar results were obtained using the filtered back-projection (FBP) technique\footnotemark[6] (see Appendix in \citealt{mar88}), confirming that the H$\beta$ MEM images are not affected by the removal of the broad absorption troughs. 

The trailed spectra show clear changes in the line behaviour: while on March 31 the He{\sc ii} $\lambda4686$ line and, to a lesser extent, the Balmer lines show a double-peaked profile with an S-wave component (which is also detectable on April 1), only enhancement in the red/blue peak of the profile is observed during the other nights. On the other hand, two emission structures are visible in the tomograms: the first is a ring, clearly visible in the Balmer images, which is a signature of emission arising from a rotating accretion disk. The second is an intense bow-shaped emission region, present both in the Balmer and He{\sc ii} $\lambda4686$ maps. Its position in the -V$_X$, +V$_Y$ quadrant places the plausible origin for this emission in the accretion gas stream, in the hotspot (the stream/disk impact region) or both. The presence of some of these physical structures on different nights argues for continuous mass transfer from the companion star during our observations.

We have plotted the theoretical path of the gas stream and the Keplerian velocities of the disk along the stream for K${_2}$=698 km s$^{-1}$ (where K${_2}$ is the radial velocity semi-amplitude of the companion star; \citealt{mcc01A}) and a mass ratio $q=0.07$ (a reasonable value for a late-type secondary star and a $7~$M$_\odot$ compact object). The Doppler maps appear to show that the region of enhanced intensity only partially overlaps the region delineated by these trajectories. A similar departure from the theoretical paths has been observed in the CVs OY Carinae \citep{har96} and WZ Sge \citep{spr98}. Note however that the accuracy of our ephemeris is such that random offsets amounting to 0.06 in orbital phase (i.e. an angular offset of about 20$^{\circ}$) are possible. If the maps are rotated clockwise by $\sim$20$^{\circ}$, the bright spot would lie between both trajectories. Again this would be a strong indicator for the presence of a hotspot \citep[see][]{mar90} or/and the accretion gas stream.  

\section{DISCUSSION}
\subsection{Disc Precession}

Asymmetric emission line profiles have been observed in several SU UMa systems, i.e. in those dwarf novae (DNe) which suffer superoutbursts \citep[for a review see][]{war95}. After maximum light, the SU UMa optical light curves exhibit a modulation at a period slightly longer than the binary orbital period, called the superhump period. A tidal resonance model (see e. g. \citealt{whi91}) is commonly used to explain the superhump period: at this stage of the superoutburst an extended accretion disk forms which, as a consequence of the tidal influence of the secondary, becomes elliptical and precesses. It is presently believed that enhanced viscous energy dissipation due to the tidal stresses from the secondary is the dominant superhump light source. Such a disk has been used to explain, for example, the asymmetric H$\alpha$ profile in OY Carinae \citep{hes92} and the systematic variations in the velocities of the central absorption feature observed in the emission lines of ZCha \citep{vog81,hon88}. Theoretical line profiles obtained using a model of a precessing eccentric accretion disk suggest that this is the case for IY UMa \citep{wu01}. Similarly the long term variations we observe in the line profiles of XTE~J1118+480 can be explained by the presence of an eccentric disk precessing around the compact primary with a period longer than the 4.1 hr orbital period. The detection of superhumps during the outburst \citep{uem00B} supports this suggestion. Indeed, our May 25 H$\beta$ map in particular suggests an elliptical disk (see Figure~\ref{fig8}). However, the extent of the bow-shaped emission region makes any such ellipticity very difficult to quantify.

An estimation of the precession period can be calculated using
$P{_{prec}}={P_{orb}}{P_{sh}}/(P{_{sh}}-P{_{orb}})$ (see
\citealt{war95}), where P$_{prec}$ and P$_{sh}$ are the precession and
superhump periods respectively. The values of P$_{sh}$ reported by
\citet{uem00B} and P$_{orb}$ (\S\ref{tomography}) imply a precession
period of $\sim30-80$~d. A precession period near quiescence of about
52~d was estimated by \citet{cas01}. Since the motion of the compact
object is expected to be small (K${_1}$=qK${_2}$=49 km s$^{-1}$), we
searched for a long term periodic modulation in the centroids of the
He{\sc ii}~$\lambda$4686 line profiles shown in Figure \ref{fig3}. We
focused on this line because its measurement was unaffected by any
nearby absorption feature. The centroid of the profiles was obtained
by fitting a gaussian after masking the core of the line. A period
search did not show any significant modulation within the expected
period range. In addition, we searched the red/blue peak flux ratio of
the HeII line for variability consistent with the predicted precession
period range, but also without success. This perhaps is not surprising
as we have only four nights of uniform orbital coverage.

\subsection{X-ray Heating vs Viscous Dissipation}

Does X-ray heating generate a significant amount of the optical flux from the disk? The slope of the optical spectrum might provide an answer, as for outbursting (optically thick) disks, standard disk theory predicts ${F_\lambda}\propto{\lambda}^{-2.33}$ if there is no X-ray heating \citep{bea84}. However, the wide range of power law indices ($-1 < \alpha < -3$) measured by \citet{sha96} in their sample of LMXBs indicates that the slope of the optical spectrum alone does not seem to be a reliable indicator of X-ray heating. On the other hand, the observation of superhumps in XTE~J1118+480 and other black hole XRNe during outburst \citep{odo96} does not in itself exclude X-ray heating as the dominant source of optical luminosity \citep{has01}. 

Another way to estimate the contribution from X-ray irradiation is to use the relation between the absolute magnitude and the orbital period/X-ray luminosity for LMXBs found by \citet{van94}. \citet{wag01} derive a distance of d=1.9$\pm$0.4~kpc for XTE~J1118+480. This implies an absolute magnitude of M${_V}$=1.6$^{+0.5}_{-0.4}$ in disagreement with the expected value for an irradiation-dominated disk of M${_V}\approx$~3.0$\pm$0.5 \citep[Fig. 2]{van94}. Here we have corrected the resultant absolute magnitude by -~0.7 mag to take into account that the visual luminosity due to X-ray reprocessing in the disk scales with the compact object mass as $\sim$~M${_{1}}^{1/3}$. A 7~M$_\odot$ black hole was assumed. Furthermore, it is clear from Fig. 2 of \citet{uem00C} that on at least two occasions, the source of the optical flux was other than X-ray irradiation. The first, before the end of the January outburst, was an optical flare (of amplitude $\gtrsim$ 1.5 mag) without any accompanying X-ray flare. In the second event, at the beginning of the March outburst, the steep rise in the optical flux was accompanied by a much slower increase in the X-ray flux. Hence we think it unlikely that X-ray reprocessing dominates the outburst optical flux.

Viscous dissipation in the disk is an alternative explanation of the
optical flux. Using the M$_V$-P$_{orb}$ relation for DNe in outburst
\citep{war87}, we have estimated M${_V}$=4.7 for a DN with an orbital
period of 4.1 hr. But since the area of the disk scales with the
primary mass as M${_{1}}^{2/3}$ for a fixed orbital period
\citep{can98} and the energy released at a given distance from the
primary scales with the primary mass as M$_1$ \citep{fra85}, we expect
the optical flux generated by a disk in a black hole XRN to be
$\sim$~M${_{1}}^{5/3}$ higher than that emitted by a disk surrounding
the $\sim$1~M$_\odot$ primary in a DN. This correction implies
M${_V}\sim1.2$ if XTE~J1118+480 contains a 7~M$_\odot$ black
hole. Note that we have assumed a moderate inclination for the binary
system. For an inclination $i=70^{\circ}$, the outburst absolute
magnitude of the viscously heated disk corrected using the
prescription of \citet{war87} is M$^{corr}_{V}\sim1.9$, close to the
observed absolute magnitude. For $i\simeq80^{\circ}$, the inclination
favoured when fitting the near quiescence optical light curve with an
ellipsoidal model \citep{mcc01A,wag01}, the above correction gives
M$^{corr}_{V}\sim2.8$. Although this magnitude is fainter than the
observed absolute magnitude, DNe superoutbursts are 0.5-1 magnitudes
brighter than normal outbursts and even 2 magnitudes brighter (or
more) in the case of the TOAD superoutbursts \citep{how95,kuu96}. If
the outburst in XTE~J1118+480 is similar to a DN superoutburst then
M$^{corr}_{V}\sim1.8$. The presence of a bright, viscously heated disk
is supported by the detection of a significant Balmer jump in the
ultraviolet spectrum \citep{has00,mcc01B}, which suggests an
appreciable thermal disk contribution to the emission. Hence the
viscously heated disk is likely to be a significant contributor to the
optical flux during the outburst of XTE~J1118+480 as in the SU UMa
systems.

\section{CONCLUSIONS}

Multi-epoch optical spectroscopic observations of XTE~J1118+480 yield
valuable information about the properties and evolution of the
accretion disk during outburst. The data reveal strong
double-peaked H$\alpha$, H$\beta$, and He{\sc ii} $\lambda$4686
emission lines whose double-peak intensity varies on time scales
longer than the 4.1 hr orbital cycle. These changes in the line
profiles can be interpreted as resulting from a precessing eccentric
disk around the compact primary. However, a search for periodic
variability in the He{\sc ii} $\lambda$4686 emission line failed to
find a modulation within the estimated precession period
range. Therefore conclusive evidence for a precessing disk in the
system is still required. Doppler tomograms display a bright
bow-shaped emission in the -V$_X$, +V$_Y$ quadrant which we interpret
as emission from the accretion gas stream, the hotspot or both. A more
accurate ephemeris is needed to properly constrain the location of the
emission within the binary system. A comparison with DNe in
superoutburst shows that viscous dissipation in the disk may make a
significant contribution to the optical luminosity during outburst.

\acknowledgements 

Use of MOLLY, DOPPLER and TRAILER routines developed largely by
T. R. Marsh is acknowledged. We thank Ann Esin for helpful comments
and Makoto Uemura for providing the VSNET light curve. We are grateful
to the anonymous referee for useful comments which improved the
quality of the manuscript. MRG was supported by NASA contract
NAS8-39073.

\clearpage

\clearpage


\clearpage


\clearpage


\clearpage

\clearpage

\clearpage

\clearpage

\clearpage
\begin{table}
\begin{center}
\caption{Journal of Observations\label{Emissiontable}}
\begin{tabular}{lccccc}
\tableline\tableline
\\
{\em Date } & {\em No. spectra} & {\em Exp. time} & 
{\em HJD start}& {\em HJD end} & {\em seeing}\\
(\em UT) & & {\em (s)} & (+2,451,000.)& (+2,451,000.) & (arcsec) \\
\\
\tableline
\\
00 March 31\tablenotemark{a}    & 15 & 1$\times$300,9$\times$600,3$\times$900,2$\times$1200  & 634.7056 & 634.9375 & 2-3 \\
00 April 1     & 7  & 600 & 635.6384 & 635.9361 &  1-2 \\
00 ~~~,,~~~2   & 3  & ,,  & 636.6492 & 636.9576 &  ,, \\
00 ~~~,,~~~3   & ,, & ,,  & 637.6169 & 637.9586 & ,, \\
00 ~~~,,~~~4   & ,, & ,,  & 638.6186 & 638.9340 & ,, \\
00 ~~~,,~~~6   & 7  & ,,  & 640.6394 & 640.9806 & 1 \\
00 ~~~,,~~~12\tablenotemark{a}  & 43 & 360 & 646.6485 & 646.8406 & 1-2 \\
00 ~~~,,~~~26  & 2  & 600 & 660.6628 & 660.8245 & ,, \\
00 ~~~,,~~~27  & 1  & ,,  & 661.6482 & 661.6482 & ,, \\
00 ~~~,,~~~29\tablenotemark{a}  & 35 & 480 & 663.6204 & 663.8294 & '' \\
00 May 8       & 1  & 300 & 672.6557 & 672.6557 & 1-2 \\
00 ~~~,,~~~10  & ,,  & ,,  & 674.7523 & 674.7523 & ,, \\
00 ~~~,,~~~11  & ,,  & 600 & 675.6582 & 675.6582 & ,, \\
00 ~~~,,~~~12  & ,,  & ,,  & 676.6679 & 676.6679 & ,, \\
00 ~~~,,~~~24  & ,,  & ,,  & 688.6886 & 688.6886 & ,, \\

00 ~~~,,~~~25\tablenotemark{a}  & 44 & 40$\times$300,4$\times$600  & 689.6596 & 689.8398 & ,, \\

00 ~~~,,~~~29  & 3  & 1$\times$60,2$\times$180 	& 693.6482 & 693.6534 & 2 \\

00 ~~~,,~~~30  & ,,  & 300 & 694.6414 & 694.6503 & ,, \\
00 ~~~,,~~~31  & 5  & ''  & 695.6403 & 695.6569 & 1-2 \\
00 July 6      & 1  & 900 & 731.6686 & 731.6686 & 1 \\
00 ~~~,,~~~10  & 8  & 300 & 735.6467 & 735.6733 & 1-2 \\
\\
\tableline
\end{tabular}
\tablenotetext{a}{Orbital phase complete or nearly complete.} 
\end{center}
\end{table}

\begin{table}
\begin{center}
\caption{Fits to the H$\alpha$ and He{\sc ii}~$\lambda4686$ Line Profiles\label{haheii}}   
\begin{tabular}{lcccccc}
\tableline\tableline
\\
	    & \multicolumn{2}{c}{\hrulefill~H$\alpha$ \hrulefill} & & \multicolumn{2}{c}{\hrulefill~He{\sc ii}~$\lambda4686$}\hrulefill& \\
{\em Date } & V$_{b}$       & V$_{r}$       & & V$_{b}$         & V$_{r}$       &\\
{\em UT}    & (km s$^{-1}$) & (km s$^{-1}$) & & (km s$^{-1}$)   & (km s$^{-1}$) &\\
\tableline					     						
March 31   &-643$\pm$24&598$\pm$20& &-881$\pm$30& 721$\pm$30&\\
April 12   &-601$\pm$29&521$\pm$17& & ---       & 707$\pm$13&\\
April 29   &-576$\pm$14&620$\pm$25& &-756$\pm$12&1226$\pm$50&\\
May 25     & 	---    &578$\pm$16& &   ---     & 760$\pm$26&\\
\tableline
\end{tabular}
\tablecomments{The table shows the shift respect to the rest wavelength of the line of the blue (V$_{b}$) and red (V$_{r}$) peaks (whenever they were measurable). The uncertainties were calculated after the error bars had been scaled to give ${\chi^2_\nu}=1$.}
\end{center}
\end{table}

\begin{table}
\begin{center}
\caption{Fits to the H$\beta$ Line Profile\label{hb}}
\begin{tabular}{lcccccccccc}
\tableline\tableline
\\
	    & \multicolumn{4}{c}{\hrulefill~3-gaussian fit \hrulefill} & & \multicolumn{3}{c}{\hrulefill~1-gaussian fit \hrulefill}& \\
{\em Date } & V$_{b}$     & V$_{r}$    & V$^{abs}$    & FWHM$^{abs}$ & & V$^{abs}$         & FWHM$^{abs}$ &EW$^{abs}$& \\
{\em UT}    & (km s$^{-1}$) & (km s$^{-1}$)& (km s$^{-1}$) & (\AA)    & & (km s$^{-1}$) &  (\AA)   &(\AA)&\\
\tableline					   			      				      
March 31 &-490$\pm$63&690$\pm$46 & 122$\pm$37 &51$\pm$3 &&168$\pm$31&63$\pm$3&2.5$\pm$0.2&\\
April 12 &-661$\pm$51&494$\pm$41 & 144$\pm$30 &49$\pm$2 &&186$\pm$21&50$\pm$2&2.6$\pm$0.2&\\
April 29 &-656$\pm$27&451$\pm$155& 373$\pm$132&36$\pm$11&&300$\pm$60&46$\pm$4&1.8$\pm$0.2&\\
May 25   &   ---     &718$\pm$21 &-120$\pm$48 &53$\pm$10&&-71$\pm$29&64$\pm$3&2.7$\pm$0.1&\\
\tableline 
\end{tabular}
\tablecomments{Columns 2-5 show the results from the 3-gaussian fit: shift of the blue (V$_{b}$) and red (V$_{r}$) emission peaks; shift (V$^{abs}$), FWHM (FWHM$^{abs}$) of the absorption component. Columns 6-8 show the results from a 1-gaussian fit after masking the emission core. Column 8 gives the area (i.e. EW) of the gaussian. Again the uncertainties were calculated after scaling the error bars to give ${\chi^2_\nu}=1$.}   \end{center}
\end{table}

\begin{table}
\begin{center}
\caption{FWZI$\tablenotemark{a}$~~and Equivalent Widths$\tablenotemark{a}$~~for H$\alpha$, H$\beta$, and He{\sc ii}~$\lambda4686$\label{fwzi}}
\begin{tabular}{lcccccccccc}
\tableline\tableline
\\
	    & \multicolumn{2}{c}{\hrulefill~H$\alpha$ \hrulefill} & & \multicolumn{3}{c}{\hrulefill~H$\beta\tablenotemark{b}$ \hrulefill} & & \multicolumn{2}{c}{\hrulefill~He{\sc ii}~$\lambda4686$}\hrulefill& \\
{\em UT Date } & FWZI & EW & & FWZI$^{abs}$& FWZI$^{em}$ &EW$^{em}$ & & FWZI & EW & \\
\tableline					   			      				   March 31    &69$\pm$11&-1.9$\pm$0.2&&95$\pm$4&49&-0.70&&53$\pm$5 &-1.2$\pm$0.2&\\
April 12    &63$\pm$2 &-1.6$\pm$0.1&&98$\pm$5&49&-0.61&&69$\pm$19&-0.8$\pm$0.3&\\
April 29    &71$\pm$6 &-2.1$\pm$0.1&&82$\pm$9&54&-0.95&&55$\pm$2 &-1.4$\pm$0.1&\\
May 25      &60$\pm$10&-0.9$\pm$0.2&&97$\pm$2&47&-0.49&&67$\pm$2 &-0.9$\pm$0.2&\\
\tableline 
\end{tabular}
\tablenotetext{a}{In angstroms.}
\tablenotetext{b}{FWZI of the absorption (FWZI$^{abs}$) and emission (FWZI$^{em}$) components of the line profile. EW$^{em}$ denotes the equivalent width of the emission component within the broad absorption. Both FWZI$^{em}$ and EW$^{em}$ are lower limits only.}
\end{center}
\end{table}

\begin{table}
\begin{center}
\caption{Line Equivalent Widths\tablenotemark{a}~~for He{\sc i} and Bowen Blend Lines\label{ew}}
\begin{tabular}{lccccccccc}
\tableline\tableline
\\

{\em Date } & He{\sc i} & He{\sc i}& He{\sc i} & Bowen  &  \\
{\em UT}    & 6678      & 5875     & 4921      & blend  &  \\
\tableline					   			      				      
March 31    &-0.6$\pm$0.1&-0.3$\pm$0.1&-0.2$\pm$0.1& ---        &\\
April 12    &-0.5$\pm$0.1&-0.5$\pm$0.1&---         & ---        &\\
April 29    &-0.6$\pm$0.1&-0.3$\pm$0.1&-0.3$\pm$0.1&-0.2$\pm$0.1&\\
May 25      &-0.3$\pm$0.1&-0.3$\pm$0.1&---         & ---        &\\
\tableline 
\end{tabular}
\tablecomments{Only the equivalent widths for the stronger lines were measured.}          
\tablenotetext{a}{In angstroms.}
\end{center}
\end{table}

\end{document}